\documentclass[preprint,11pt]{elsarticle}

\usepackage{amsmath}
\usepackage{subcaption}
\usepackage{float}
\usepackage{color}
\usepackage{framed,multirow}

\usepackage{amssymb}
\usepackage{latexsym}

\usepackage{url}
\usepackage{xcolor}
\usepackage{fullpage}
\usepackage{graphicx}
\usepackage{float}

\usepackage{subcaption}
\usepackage{hyperref}
\definecolor{newcolor}{rgb}{.8,.349,.1}

\journal{Journal of Computational Physics}

\begin{document}

\begin{frontmatter}

%
%
%

\title{Efficient numerical algorithm for multi-level ionization of high-atomic-number gases}

\author[1]{Aiqi Cheng}
\author[1,2]{Roman Samulyak\corref{cor1}}
\cortext[cor1]{Corresponding author at: Department of Applied Mathematics and Statistics, Stony Brook University, Stony Brook, NY 11794, United States \\
\textit{Email:} \url{roman.samulyak@stonybrook.edu}}


\address[1]{Department of Applied Mathematics and Statistics, Stony Brook University, Stony Brook, NY 11794, United States}
\address[2]{Computational Science Initiative, Brookhaven National Laboratory, Upton, NY 11973}

\begin{abstract}
An efficient numerical algorithm for the laser driven multi-level ionization of high-atomic-number gases is proposed and implemented in an electromagnetic particle-in-cell code SPACE. The algorithm is based on analytical solutions to the system of differential equations describing ionization evolution. Using analytical solutions resolves the multiscale issue of ionization due to different characteristic time scales of ionization processes and the main code time step. Algorithm efficiency is improved by using a locally-reduced system of differential equations. The effects of the orbital quantum numbers and their projections have been examined. The algorithm is applied to the study of ionization injection of electrons into laser-driven plasma wakefields.
\end{abstract}


\end{frontmatter}



\section{Introduction}

An accurate numerical model for ionization of gases by high-power lasers is critical for numerical simulations of laser-plasma interactions and for the
laser-plasma wakefield acceleration (LPWA) of particles in particular\cite{Tajima79,Esarey09,Gonsalves19}. Although modern lasers used for LPWA are capable of  completely ionizing hydrogen gas during several periods of laser filed oscillation and within a small fraction of the laser pulse duration, 
the assumption of a pre-ionized plasma is not accurate for laser intensities close to the tunneling ionization threshold
\cite{Bruhwller03,Kumar19}.
Ionization effects such as creation of plasma through tunneling ionization, frequency up-shifting of the laser, harmonic
generation, and scattering instabilities play a major role in the laser wakefield acceleration process. 
 An accurate modeling of the laser ionization of gases significantly improves the resolution of such processes.

The large interest for  multi-level ionization of high-atomic-number gases in the context of LPWA is driven by ideas of the ionization injection of electrons into plasma wakes \cite{Yu2014,Schroeder2014,ChenMin2006}, a technique that may significantly reduce the complexity of the LPWA method and improve the quality of resulting accelerated electron beams.
While the ionization of hydrogen, a typical medium for LPWA experiments, is easy to resolve numerically, multi-level ionization of high-atomic-number gases presents a set of more complex modeling challenges. These include the multiscale nature of ionization, complex atomic states characterized by sets of orbital quantum numbers and their projections, and various software development issues such as a significant increase of the memory usage if the evolution of multiple ionization states is tracked at every computational mesh block in the simulation domain. In this paper, we propose a numerical alorithm for electromagnatic Particle-in-Cell (PIC) codes that improves the accuracy and numerical efficiency compared to previously published models \cite{Chen2013}. PIC \cite{Birdsall91,Hockney88} is a common method for the simulation of laser-plasma interactions and LPWA which solves the Vlason-Maxwell equations using a hybrid particle-grid method. In PIC, the electromagnetic Maxwell equations are discretized and updated on a grid while the distribution functions of charged particles is sampled by a set of Lagrangian particles, the motion of which is coupled to the mesh via currents.

The algorithm has been implemented in SPACE, a parallel, relativistic, three-dimensional particle-in-cell code developed for the simulation of electromagnetic fields, relativistic particle beams, and plasmas \cite{YuSam22}. A distinct feature of SPACE is a software module for atomic physics transformations.

This paper is organized as follows. In Section 2, we present the main equations governing laser tunneling ionization of high-atomic-number gases. Section 3 describes details of the mathematical model and presents verification tests. We conclude this paper by Section 4 with a summary of our results and plans for the future work.

\section{Laser ionization of high-Z gases}

The Ammosov–Delone–Krainov (ADK) model \cite{Ammosov1986,Chen2013} for tunneling ionization in high-atomic-number gasses derives the following expression for the probability of ionization
\begin{eqnarray}
	W_{lm} =&&\omega_\alpha \sqrt{\frac{3{n^*}^3 E_L}{\pi Q^3 E_a}} \frac{Q^2}{2 {n^*}^2} \left(\frac{2e}{n^*}\right)^{2n^*}\exp{\left[-\frac{2E_a}{3E_L} \left(\frac{Q}{n^*}\right)^3\right]}\times
	\nonumber \\		
	&&\frac{(2l+1)(l+|m|)!}{2\pi n^* 2^{|m|} (|m|)!(l-|m|)!} \left(2\frac{E_a Q^3}{E_L {n^*}^3}\right)^{2 n^*-|m|-1}.
\end{eqnarray}
Here $\omega_a = \alpha^3 c/r_e =4.13\times10^{16} s^{-1}$ is the atomic unit frequency, $E_a = 510\ GV/m$ is the atomic unit of Electric field, and $E_L\ (GV/m)$ represents the strength of the local laser electric field. $l$ and $m$ are the electron's orbital quantum number and projection of the orbital quantum number respectively, $n^* = Q \sqrt{U_H/U_{ion}}$ is the effective principle quantum number, and $U_H$ and $U_{ion}$ are the ionization potential of hydrogen and ion of the material being ionized, respectively. $Q$ is the charge number after ionization.

Denoting the number density of atoms in the neutral ground state as $n_0$ and the number density of ions ionized to 
the state $j+$ as $n^{j+}$, the system of equation governing the ionization dynamics is as follows
\begin{align}
\label{eq:ADK_evol}
	\frac{dn_0}{dt} &= -W_0 * n_0(t), \nonumber\\
	\frac{dn^+}{dt} &= W_0 * n_0(t) - W_1 * n^+(t), \nonumber\\
	\frac{dn^{(Z-1)+}}{dt} &= W_{Z-2}*n^{(Z-2)+}(t) - W_{Z-1}*n^{(Z-1)+}(t), \\
	\frac{dn^{Z+}}{dt} &= W_{Z-1}*n^{(Z-1)+}, \nonumber
\end{align}
where $Z$ denotes the gas atomic number. 
We would like to note that because overall time scales of  LPWA simulations are short compared to recombination processes in gases, the recombination terms 
were not included in the governing system of equations.

This system of equations must be solved independently at each numerical grid cell of the computational domain which presents a set of challenging numerical issues. First, the system includes multiple time scales associated with short characteristic times for ionization dynamics of each individual level and the overall characteristic time of the electromagnetic problem being studied. Second, recording the evolution dynamics of all individual ionization levels in each computational mesh cell requires a significant increase of the total memory. 
In the next Section, we propose an efficient numerical algorithm for multi-level ionization of high-atomic-number gases that resolves these and other issues, as well as a set of verification tests for the algorithm.  

\section{Numerical model and its verification}

We start with writing equations (\ref{eq:ADK_evol}) in the following matrix form:
\begin{equation}
	\begin{bmatrix} \dot{n_0} \\ \dot{n}^+ \\ \dot{n}^{2+} \\ ... \\ \dot{n}^{(Z-1)+} \\ \dot{n}^{Z+} \end{bmatrix}
	= 
	\begin{bmatrix}
		-W_0 & 0    & 0   & ... & 0 & 0 \\
		W_0  & -W_1 & 0   & ... & 0 & 0 \\
		0    & W_1  & -W_2& ... & 0 & 0 \\
		\vdots & ... & ... & \ddots & \vdots & \vdots \\
		\vdots & ... & ... & W_{Z-2} & -W_{Z-1} & 0 \\
		0 & ... & ... & 0 & W_{Z-1} & 0
	\end{bmatrix}
	\begin{bmatrix} n_0 \\ n^+ \\ n^{2+} \\ ... \\ n^{(Z-1)+} \\ n^{Z+} \end{bmatrix}
	\label{eq:matrix_form}
\end{equation}
Since the ionization probabilities contained in the matrix depend on the local electric field values, 
our model is based on the natural assumption that ionization probabilities do not change within each time step of the 
overall electromagnetic simulation which updates states of the electromagnetic field. Assuming that the ionization probabilities are updated at the beginning of each time step of the overall simulation denoted at $t$, we compute the analytical solution to (\ref{eq:matrix_form}) at the end of the time step $t+dt$
in terms of the eigenvalues $\lambda_i = -W_i(t)$ and eigenvectors $\vec{V}_j$ of the corresponding matrix. 
The general solution is:
\begin{equation}
n^{i+} (t+dt) = \sum\limits_{j=0}^Z C_j \vec{V}_j(i) e^{-W_j(t) dt}
\end{equation}
The initial conditions at each time step $n^{j+}(t)$, $j = 0, \ldots, Z$ are used to compute coefficients $C_j$.
\begin{equation}
 	\begin{bmatrix}
 	 \vec{V}_0 & | & \vec{V}_1 & ... & | & \vec{V}_Z
 	\end{bmatrix}
 	\begin{bmatrix}
 	C_0 \\ C_1 \\ \vdots \\ C_Z
 	\end{bmatrix}
 	=
 	\begin{bmatrix}
 	n_0(t) \\ n^+(t) \\ \vdots \\ n^{Z+}(t)
 	\end{bmatrix}
\end{equation}
Analytic solutions are computed independently in every computational cell and the population of ionization levels $n^{i+}$ is updated. Using analytical solution efficiently resolves the multiscale nature of ionization evolution and it eliminates the need for fractional time stepping for a numerical solver for the ODE system  (\ref{eq:ADK_evol}), as implemented in some electromagnetic codes. Explicit expressions for the analytical solutions subject to another simplifying approximations that reduces the total storage will be presented later in this Section.

\begin{figure}[H]	
	\centering
	\includegraphics[width=0.8\linewidth]{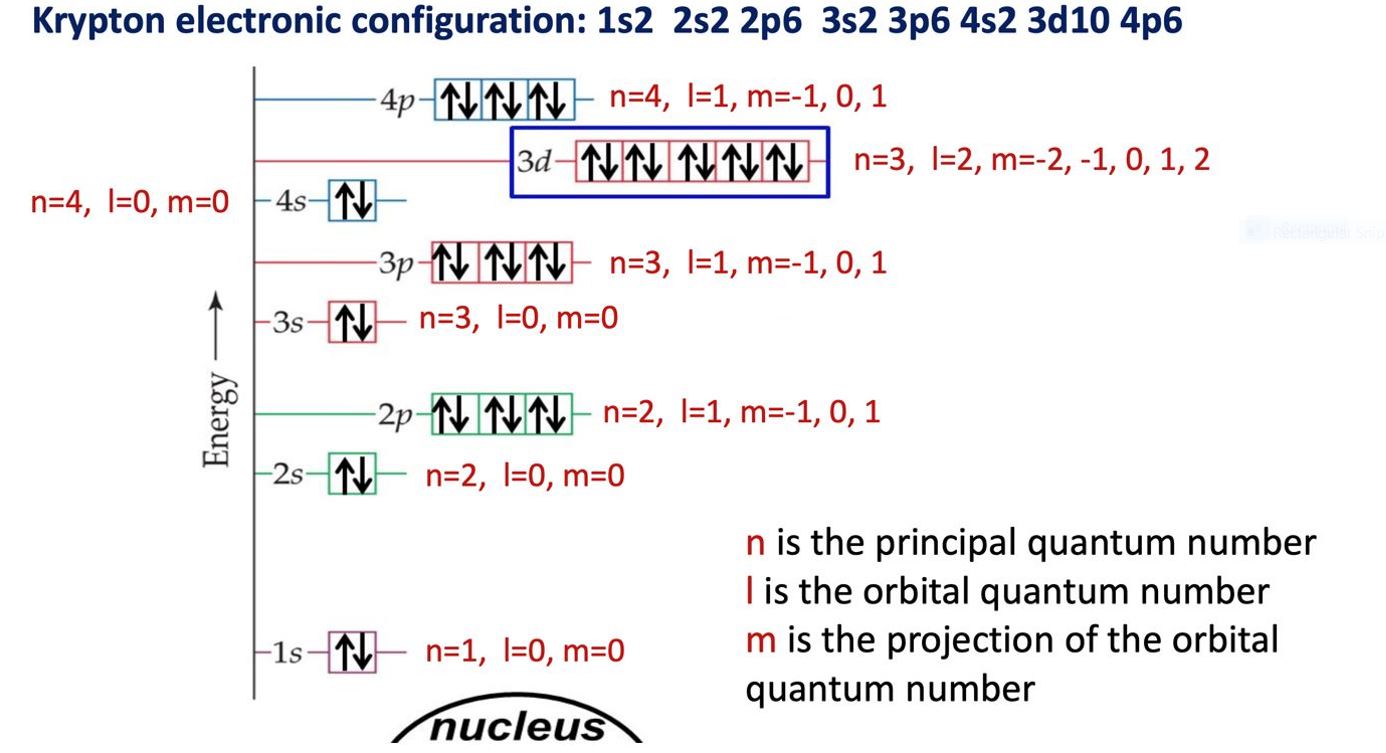}	
	\caption{Schematic the electron structure of krypton}
	\label{Kr_e_structure}
\end{figure}

The second challenge is associated with strong dependence of ionization probabilities on the orbital quantum numbers and their projections for electrons being ionized. A schematic of the electron structure of krypton, the medium of interest for the two-color injection in laser plasma wakefield acceleration, is shown in Figure \ref{Kr_e_structure}. 

In the present implementation, we assume that electrons are ionized starting from out-most orbitals and their ionization probabilities are computed according to the quantum numbers and their projections as indicated in Figure \ref{Kr_e_structure}. However, we also investigated
other scenarios in which after removing the first two electrons from the $4p$ orbital, the unoccupied p-orbitals coupled to produce five-fold degenerate singlet-$d$, nondegenerate singlet-$s$, and triplet-$s$ states, and similar transformations in the lower orbitals. Since we observed negligibly small
changes in the overall ionization dynamics, we compute probabilities based on quantum numbers presented in
Figure \ref{Kr_e_structure}.   

The final challenge is associated with a significant increase of memory allocation for computing ionization processes. To store all ionization levels in krypton, 36 floating point numbers must be added to each computational cell. Of course, the total number of ionization levels could be reduced if ionization beyond some $N+$, $N<Z$, is 
unlikely at given laser parameters. This, however, would require a preliminary study for each new simulation setting, and fine-tuning of the ionization model for every gaseous material involved in the laser-matter interaction. 

\begin{figure}[H]
	\centering
	\begin{subfigure}{0.9\linewidth}
		\centering
		\includegraphics[width=\linewidth]{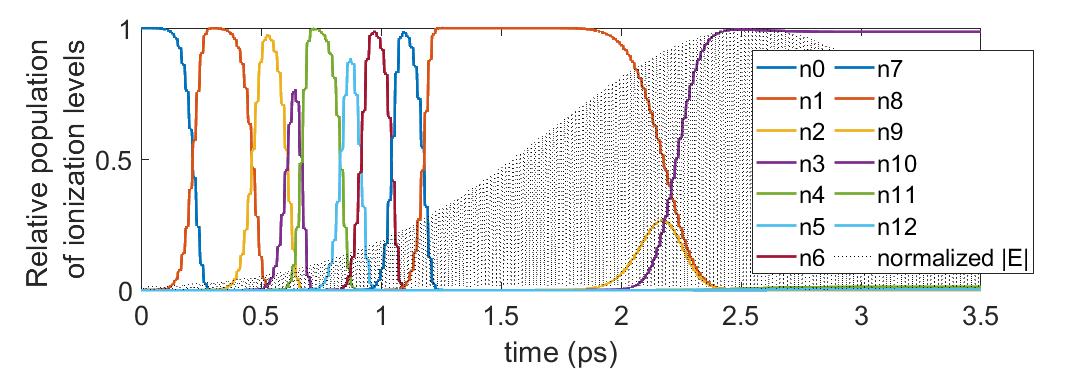}
		\caption{Full 36-level system of equations}
		\label{ionization36level}
	\end{subfigure} 
	\begin{subfigure}{0.9\linewidth}
	\centering
	\includegraphics[width=\linewidth]{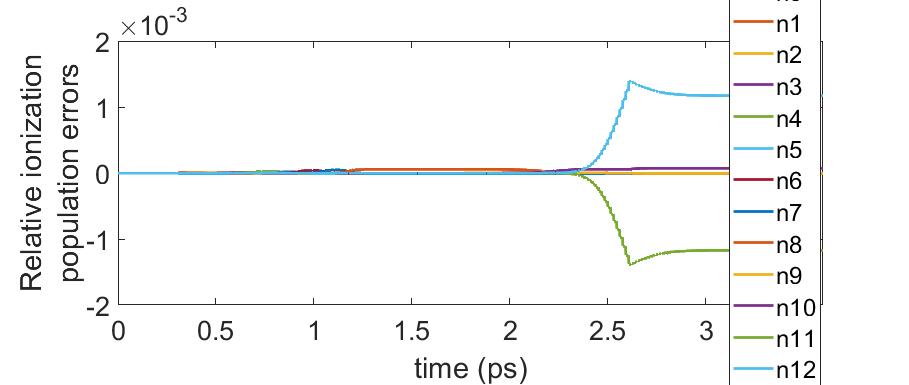}
	\caption{Errors for 4-level system of equations}
	\label{ionization_error}
	\end{subfigure} 
	\begin{subfigure}{0.8\linewidth}
	\centering
	\includegraphics[width=0.6\linewidth]{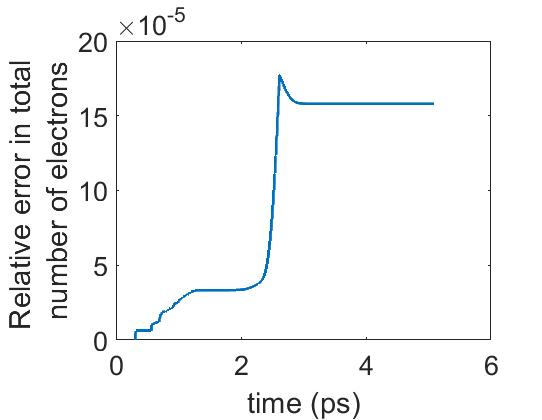}
	\caption{Relative error of the electron density}
	\label{error_e_density}
	\end{subfigure} 
	\caption{(a) Evolution of the relative population of ionization levels along the laser axis computed by analytical solutions to the full system of ODE's (\ref{eq:matrix_form}) containing 36 ionization levels.  (b) Errors of the relative population of ionization levels for solutions with a reduced 4-level system of equations. (c) Relative errors of the electron density for solutions with a reduced 4-level system of equations. A 4J, 2 ps $CO_2$ laser pulse propagated in krypton gas with the neutral density of 2.e17 1/cc was used.}
	\label{comp_4_vs_full_ionization}
\end{figure}

We performed detailed studies of the ionization dynamics of various ionization levels for different laser settings and concluded that only 2-3 intermediate levels are populated at any moment of time, with the population of lower and higher levels effectively approaching zero. Figure \ref{ionization36level} shows the evolution of ionization levels in krypton interacting with a $CO_2$ laser pulse. A 4J laser pulse with 2ps duration was used and the laser wavelength was 9.2$\mu m$. 
The data was collected from a point on the laser propagation axis. The highly oscillatory gray line shows, in normalized units, the laser amplitude evolution at the given point. We observe that the density of neutrals reduces to zero within 0.35 ps after the laser pulse arrival. Due to a high ionization probability, this happens at a low laser field amplitude. The density of $Kr+$ rapidly grows and then reduces to zero at about 0.5 ps. Higher ionization levels undergo a similar dynamics. As the laser amplitude reduces from its maximum after 2.5 ps, the population of the ionization level $Kr^{10+}$ slowly reduces and the levels $Kr^{11+}$ and $Kr^{12+}$ slowly grow but their population remains at the order of magnitude of 1\%. Similar dynamics occurs in other locations of the computational domain, off the laser axis, except that the maximum ionization level is lower due to lower laser field intensities. We observe that at any time, the effectively non-zero population is present for only 2-3 ionization levels.

This observation allowed us to replace the full system of equations for all ionization levels by a reduced-order system containing only 3 or 4 ionization levels which is specified independently for each computational cell. The active ionization levels are shifted to the higher ones independently in each computational cell when the population of the lowest level drops below a prescribed threshold.  Figure \ref{comp_4_vs_full_ionization} presents a comparison 
of the evolution of ionization levels along the laser axis computed by analytic solutions to the full system of ODE's (\ref{eq:matrix_form}). The reduced 4-level system of equations describes this process very accurately. We do not plot the corresponding solution obtained with the  4-level system since there is no visible difference with the plot \ref{ionization36level} obtained with the full system of ODE's. Instead, we plot the evolution of solution errors for the reduced 4-level system in Figure  \ref{ionization_error} and observe that errors are of the order $10^{-4}$ for the intermediate states and for the final dominant ionization state  $Kr^{10+}$. Errors of the partially ionized states  $Kr^{11+}$ and  $Kr^{12+}$ are on the order of $10^{-3}$, but these states contribute very small quantity of electrons to the final superposition state. Since errors in the final ionization states partially compensate each other, the relative error of the electron density,
depicted in Figure \ref{error_e_density}, demonstrates high overall accuracy of the method.

The analytical solution for the reduced 4-level system is:
\begin{align}
	n_p(t+dt) &= n_p(t) * e^{-W_p*dt} \\	
	n_{p+1}(t+dt) &= n_p(t) *\frac{W_p}{W_p-W_{p+1}} * (e^{-W_{p+1} dt}-e^{-W_p dt}) + n_{p+1} (t) * e^{W_{p+1} dt} \\	
	n_{p+2}(t+dt) &= n_p(t) * \left( \frac{W_p W_{p+1} e^{-W_{p+2} dt}}{(W_p-W_{p+2})(W_{p+1}-W_{p+2})} - \frac{W_p W_{p+1} e^{-W_{p+1} dt}}{(W_p-W_{p+1})(W_{p+1}-W_{p+2})} \nonumber \right.\\
	& \left. + \frac{W_p W_{p+1} e^{-W_p dt}}{(W_p-W_{p+1})(W_p-W_{p+2})}  \right) \nonumber \\
	&+ n_{p+1}(t) * \left( \frac{(W_p W_{p+1} - W_{p+1} W_{p+2}) e^{-W_{p+2} dt}}{(W_p-W_{p+2})(W_{p+1}-W_{p+2})} - \frac{(W_p W_{p+1} - W_{p+1}^2) e^{-W_{p+1} dt}}{(W_p-W_{p+1})(W_{p+1}-W_{p+2})} \right) \nonumber \\
	&+ n_{p+2}(t) * \left(\frac{(W_{p+2}^2 + W_p W_{p+1} - W_p W_{p+2} - W_{p+1} W_{p+2}) e^{-W_{p+2} dt}}{(W_p-W_{p+2})(W_{p+1}-W_{p+2})} \right) \\	
	n_{p+3} &= n_p(t) * \left( 1 + \frac{W_p W_{p+1} e^{-W_{p+2} dt}}{(W_p-W_{p+2})(W_{p+1}-W_{p+2})} + \frac{W_p W_{p+2} e^{-W_{p+1} dt}}{(W_p-W_{p+1})(W_{p+1}-W_{p+2})} \nonumber \right. \\& \left. - \frac{W_{p+1} W_{p+2} e^{-W_p dt}}{(W_p-W_{p+1})(W_p-W_{p+2})}  \right) \nonumber \\
	&+ n_{p+1}(t) * \left( 1 + \frac{(W_p W_{p+1} - W_{p+1} W_{p+2}) e^{-W_{p+2} dt}}{(W_p-W_{p+2})(W_{p+1}-W_{p+2})} + \frac{(W_p W_{p+2} - W_{p+1} W_{p+2}) e^{-W_{p+1} dt}}{(W_p-W_{p+1})(W_{p+1}-W_{p+2})} \right) \nonumber \\
	&+ n_{p+2}(t) * \left( 1 + \frac{(W_{p+2}^2 + W_p W_{p+1} - W_p W_{p+2} - W_{p+1} W_{p+2}) e^{-W_{p+2} dt}}{(W_p-W_{p+2})(W_{p+1}-W_{p+2})} \right) + n_{p+3}(t),
\end{align}
where $p$ denoted the lowest ionization level of the system. Initially, $p=0$ uniformly in the computational domain, and the value of $p$ increases non-uniformly after the cut-off criteria is met in a specific computational cell. 
 
The corresponding reduced 3-level system of ODE's has also been tested. Since numerical simulations with the 3-level system showed increased errors in certain parts of the computational domain, the 4-level system was selected as the most optimal solution that optimizes the accuracy and numerical cost.

\section{Implementation in SPACE and numerical examples}

The numerical algorithm for multi-level ionization of high-atomic-number gases was implemented as a module in SPACE \cite{YuSam22}, a parallel, relativistic, three-dimensional particle-in-cell code developed for the simulation of electromagnetic fields, relativistic particle beams, and plasmas. The algorithm proposed in this paper extends a set of atomic physics algorithms previously implemented in SPACE that resolve single-electron ionization, recombination, and electron attachment to dopants in dense neutral gases. These algorithms were used for longer-time-scale processes compared to LPWA and can be illustrated by SPACE simulations, performed in support of the experimental program on the high-pressure hydrogen gas-filled radio frequency cavity in the Mucool Test Area at Fermilab \cite{Yu2017,Yu2018}. The previous SPACE implementation also employed the ADK ionization model for hydrogen use for the simulation of LPWA 
driven by a CO2 laser \cite{Kumar2019,Kumar2021}.
 
\begin{figure}
	\centering
	\includegraphics[width=1.0\linewidth]{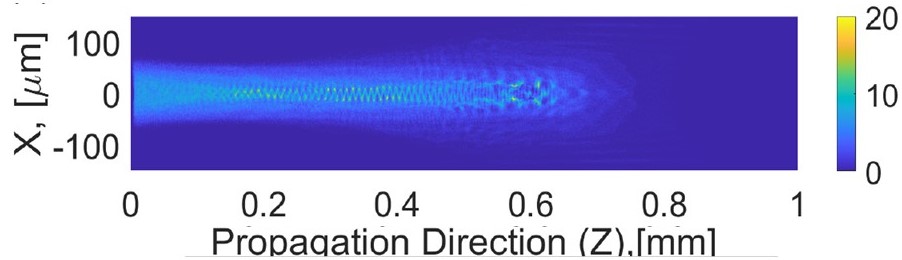}	
	\caption{3D SPACE imulation of the 4J, 2 ps $CO_2$ laser interacting with krypton gas with neutral density of $2\times 10^{17}$1/cc. Distribution of electrons on a 2D plane along the laser propagation and polarization directions is shown.}
	\label{simul_3D}
\end{figure}

The new multi-level ionization model enabled simulation studies of ionization injection of electrons into wakes created by a $CO_2$ laser pulse in high-atomic-number materials such as krypton. After the plasma wakes are excited by the $CO_2$ laser, a separate high-intensity near-infrared laser with a low normalized vector potential will inject electrons through ionization inside the wakes. Detailed simulation study of the interaction of both lasers with krypton and ionization injection of electrons will be published in a forthcoming paper. Here we illustrate the use of the multi-level ionization algorithm by the simulation of a single 4J, 2 ps $CO_2$ laser pulse creating plasma wakes in krypton gas with neutral density of $2\times 10^{17}$1/cc. Figure \ref{simul_3D} shows a distribution of electrons on a 2D plane along the laser propagation and polarization directions.

\section{Conclusions and future work}
 
In this paper, we developed an efficient multi-level laser tunneling ionization algorithm in high-atomic-number gases. The algorithm features several improvements compared to the previously published methods based on numerical solutions to the system of ODE's describing ionization dynamics via fractional time stepping. By using analytical solutions to the system of differential equations describing ionization evolution, our method effectively resolves the multiscale nature of ionization processes which occur at time scales significantly different compared to the time step of the main code. The method also eliminates the need to store all ionization states in every cell of the numerical mesh by using a locally-reduced system of equations, thus significantly reducing the memory requirement. It also accounts for electron quantum numbers and their projections while computing ionization probabilities. 

The multi-level ionization algorithm was implemented and verified in the code SPACE. The algorithm developed in this paper extends a set of atomic physics algorithms previously implemented in SPACE that resolved single-level ionization, electron-ion recombination, and electron attachment to dopants in dense neutral gases. 
SPACE has been extensively used in a variety of applications in the area of plasma physics and accelerator design, in particular for simulations of coherent electron cooling of relativistic ion beams \cite{Ma2018} and the laser-plasma wakefield acceleration \cite{Kumar2019,Kumar2021}. The multi-level ionization module is currently being applied to numerical studies of the ionization injection and acceleration of electrons into plasma wakes within the Brookhaven National Laboratory program on 
$CO_2$ laser driven plasma wakefield accelerator.

\section*{Acknowledgments}
This work was supported by Grant No. DE-SC0014043 funded
by the U.S. Department of Energy, Office of Science, High Energy
Physics.

%


\end{document}